\documentclass[a4paper]{jpconf}
\usepackage{graphicx}
\usepackage{array}
\usepackage{url}
\usepackage{float} 
\usepackage{xcolor}
\usepackage{wrapfig} % per scorrimento testo

\begin{document}
\title{Exploring active learning in physics with ISLE-based modules in high school}

\author{E. Tufino$^{1}$ , P. Onorato$^{1}$ and S. Oss$^{1}$}

\address{$^1$ Department of Physics, University of Trento, 38123 Trento, Italy}

\ead{eugenio.tufino@unitn.it}

\begin{abstract}
This study presents a case study of active learning within the Investigative Science Learning Environment (ISLE), using the iOLab digital devices. We designed a pilot lab format to enhance student engagement and understanding through direct experimentation, taking advantage of the multifunctional capabilities of the iOLab devices. This paper evaluates the pedagogical effectiveness of integrating ISLE with digital tools for data collection and analysis in physics experiments. The initial findings provide insights into the pedagogical benefits and logistical considerations of using such technologies in a laboratory setting. Although no direct comparison with traditional teaching methods has been made, the observed student engagement and feedback suggest a positive impact on learning outcomes, even within the constraints of the short duration of the interventions.

\end{abstract}

\section{Introduction}
\label{Introduction}
The introduction of active learning strategies is important in both physics lectures classrooms and in laboratory courses. Despite lab activities could be seen as interactive as students engage with various equipment and each other in group work, recent research indicates that traditional confirmation-style labs, at first year undergraduate level, characterized by step-by-step instructions and pre-determined outcomes, are ineffective in terms of both content learning \cite{Wieman2015} and development of lab skills, as well as in fostering positive student attitudes towards laboratory work \cite{Wilcox2016}. Possible intervention strategies thus involve reducing the detailed instruction to be
given to students to leave room for their decision-making and limiting the confirmatory aspects of lab activities \cite{Holmes_Smith21}. 

A review of alternative pedagogies adopted by physics educators in laboratory settings reveals a variety of innovative approaches such as SCALE-UP, Real-Time Physics (RTP) \cite{Sokoloff}, Modeling Instruction \cite{Brewe_Modeling}, ISLE (Investigative Science Learning Environment) \cite{Etkina2015a}, and others, each with unique characteristics. Particularly at the high school level, the Modeling Instruction Framework and ISLE, the latter developed by Etkina and her collaborators, stand out for their broad applicability and versatility. Our research focuses primarily on the implementation of ISLE because not only does this methodology place students at the centre of the learning process, but it also emphasises growth mindset and formative assessment. ISLE's approach, which includes the use of rubrics to assess scientific abilities, is consistent with our focus on fostering a deep understanding of physics through active, inquiry-based learning. In fact, ISLE explicitly teaches students to generate and test alternative hypotheses to explain a phenomenon. Research has validated ISLE's effectiveness in both high school and undergraduate settings, making it an exemplary framework for developing the critical thinking and scientific inquiry skills essential for future citizens. The validation and dissemination of ISLE have primarily taken place in the US, but recent efforts with the ISLE approach have also been made in countries such as Slovenia \cite{Faletic2020} and Sweden \cite{Samuelsson2022}, which are characterised by different contexts. This work contributes to the exploration of ISLE in Italy \cite{Bologna2024}.

In addition, the emergence of hands-on technological tools, ranging from smartphone sensor apps to Arduino microprocessors, has the potential to significantly enhance inquiry-based approaches in the laboratory, particularly in the development of skills related to data analysis and laboratory practice \cite{organtini2022}. In this work, we have primarily used the iOLab device, developed by the PER group at the University of Illinois \cite{iolab}, as a pedagogical tool for teaching the fundamentals of data analysis and active learning. It allows students to carry out laboratory exercises across the full range of high school Physics and beyond, whether at home or in the classroom, using a single device and reducing cognitive load \cite{Sweller2020}. In the following sections, we describe the laboratory modules based on the ISLE approach, using iOLab device.

\section{Our integrated approach: ISLE methodology and new digital technologies}
ISLE encourages students to engage in a process of knowledge construction that mirrors professional scientific practices \cite{Etkina2015a}. The process consists of generating explanations in groups, testing them through experiments, and refining them to create a dynamic and exploratory learning environment (for an illustration of the ISLE process, see Figure 8 in \cite{Etkina2015a}).
ISLE focuses on three types of experiments: observational, testing and application. In observational experiments, students observe phenomena proposed and carefully selected by the teacher and collect data in groups to develop explanations. Testing experiments are devised by students to assess their own explanations. In doing so, they must employ hypothetical-deductive reasoning skills and try to find out which of their explanations are falsified by the experiments. Finally, application experiments allow students to apply their knowledge and skills to real-world situations, solving a realistic problem or determining an unknown quantity.

The emphasis on testing and investigation is fundamental to the ISLE approach.This process not only deepens their understanding of physical concepts, but also promotes the development of scientific thinking and problem solving skills. Collaborative learning is a key component of the ISLE methodology. By working in groups, students improve their communication skills, which are essential for real scientific work. Another key intention of the ISLE methodology, as highlighted in \cite{brookes2020}, is the emphasis on student well-being. This approach emphasises the importance of fostering a growth mindset in students - a psychological perspective that views intelligence and ability as qualities that can be developed through effort, practice and perseverance, as explored in the literature \cite{grow_mindset}.

ISLE is a form of 'guided' inquiry, where 'guided' is used in the sense specified by Buck et al.\cite{Buck2008} in their work on inquiry-based laboratory education. In their study they classify the different levels of inquiry, with higher levels corresponding to progressively less structure provided by the teacher, a concept further supported by the work of Holmes and Smith cited above.

ISLE has been effectively implemented in many US high schools \cite{Buggé_longterm}, demonstrating a close alignment with the NGSS standards \cite{NGSS}, which are widely adopted in these institutions.

The adaptation of ISLE to the Italian educational and cultural context required addressing specific challenges, as highlighted by Bologna et al. \cite{Bologna2022}. An additional consideration, in our case, is the limited class time available for laboratory activities and the fact that our interventions had to be scheduled after the theoretical concepts had first been introduced in the classroom by the students' teachers, which differs from the ideal ISLE model in which theory and practice are more closely intertwined. This adaptation reflects a wider challenge within the Italian educational system to accommodate active learning approaches within the constraints of existing timetables and curricular structures.

The introduction of digital technologies, such as Arduino, smartphones with dedicated applications and sensor devices like iOLab \cite{iolab}, is significantly changing pedagogical practices in physics education. These tools add variety and innovation to the laboratory experience, fostering a more interactive and immersive learning environment. We specifically chose the iOLab system  for its consistent and reliable sensors and its adaptability to a wide range of experiments thanks to its force sensor and voltage pins  (although other viable devices exist on the market). This choice addresses the issue of the different sensor capabilities of current smartphones. It also avoids the legal risks associated with using students' personal devices. In addition, the use of a single device and intuitive graphical interface for real-time data analysis allows students to focus on understanding and applying physics concepts without the distraction of managing multiple, disparate pieces of equipment. 

\subsection{Implementing Learning Goals}
\label{learning_goals}

Our instructional design, based on AAPT guidelines \cite{AAPT2014} and NGSS practices \cite{NGSS} and inspired by the formulation of categories in Bloom's Revised Taxonomy \cite{Bloom_revised}, emphasises three fundamental learning goals.

\begin{itemize}

\item Conceptual Knowledge: For this learning goal, we focused specifically on three levels of Bloom's Revised Taxonomy: Remembering, Understanding, and Applying, which are appropriate for high school students. With regard to Remembering, students will be able to recall the physical concepts underpinning the proposed experiments. For Understanding, they will be able to explain, compare, and interpret these concepts. Finally, for Applying, they will be able to use this knowledge to solve simple experimental problems.

\item Technical skills (in relation to iOLab): Students should be able to use the iOLab device and the phyphox application with smartphones, including setup, data collection and analysis, and problem solving.
\item ISLE process: Students should be able to apply the ISLE methodology and development of scientific abilities \cite{Etkina2006Abilities}. Specifically, they should formulate multiple explanations for the observed patterns. They should test these by designing experiments, collecting, and representing data. Finally, they should be able to apply the knowledge gained to determine the value of some physical quantities. 
\end{itemize}

\subsection{Overview of the modules}
In line with the above learning goals, the proposed modules, of 4 to 10 hours, aim to progressively develop students' familiarity with experimental physics techniques, conceptual knowledge, and inquiry based approach. Beginning with an introduction to the iOLab instrument, students perform basic measurement tasks and progress to more complex ISLE-based activities using the instrument alongside basic laboratory physics materials. 

The syllabus of the modules has been designed to include some review questions to be solved in groups to consolidate the concepts and skills acquired. This approach not only improves students' experimental and analytical skills, but also introduces a growth mindset through deliberate practice \cite{deliberate}, including immediate feedback and explicit goals. Table \ref{tab:overview} gives an overview of the modules implemented in some 11th and 12th grade classes at the High School, known in Italy as "Liceo Scientifico" and "Liceo Scientifico opzione scienze applicate"
\begin{table}[htbp]
\caption{\label{tabFirstSem}Overview of the physics modules including classroom implementation. Resources for lab activities can be found in the GitHub repository \cite{GithubRep}.}
\label{tab:overview}
\centering
\renewcommand{\arraystretch}{1.5}
\begin{tabular}{p{3cm} p{5cm} p{5cm}}
\hline
Module & Lab Activities & Classroom Implementation \\
\hline
Mechanics & Dynamic friction, Horizontal Atwood machine, circular motion & 3 classrooms, 11th grade \\
Electric Field and Electric Potential & Electric potential & 2 classrooms, 12th grade \\
DC Circuits & Light bulb/LED circuits, Ohm's law, LED I-V characteristic curve & 2 classrooms, 12th grade \\
\hline
\end{tabular}
\end{table}
From the overview presented in the Table \ref{tab:overview}, it is important to recognize that, although the specific contexts and applications differ across the modules, the overarching objectives in developing key scientific skills remain consistent. These include developing and testing multiple explanations and mathematical relationships, proficiently using scientific equipment, collecting and analyzing data, evaluating the outcomes of experiments, ability to apply the knowledge gained in estimating physical quantities. 

\subsection{Newton's Laws Applications for 11th grade students}
\begin{figure}
\begin{center}
\includegraphics[width=6in]{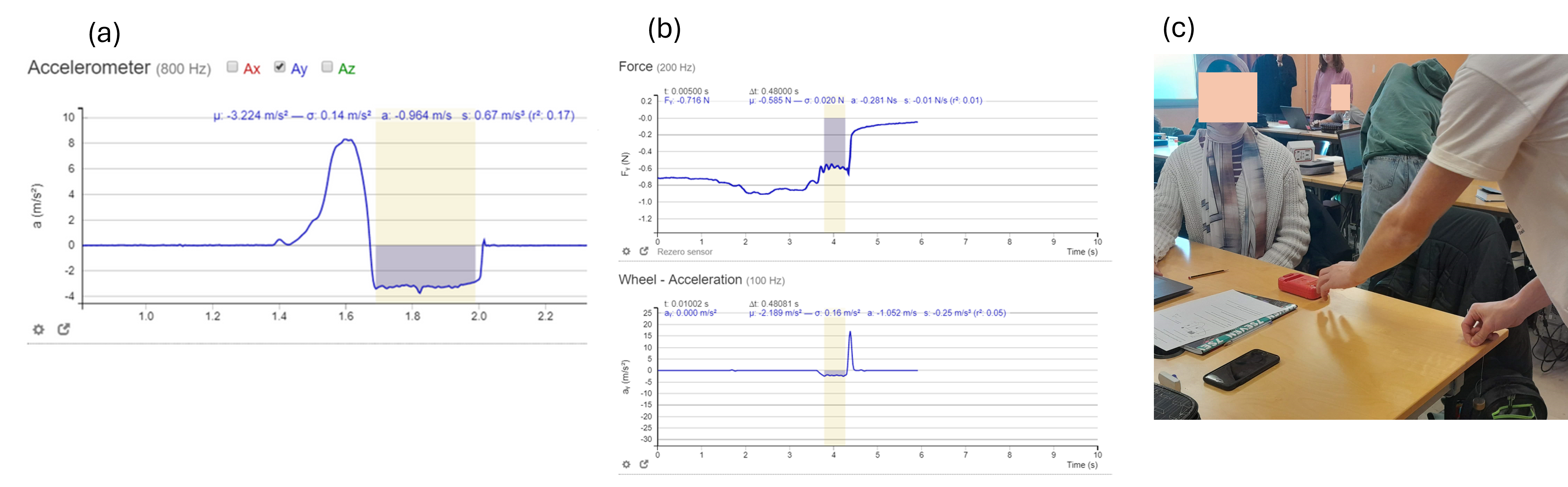}
\end{center}
\caption{\label{fig:mechanics} (a) Acceleration measured using the iOLab device sliding on a table. The highlighted area corresponds to the segment where only static friction is present ($\mu$ represents the average acceleration in this segment, $\sigma$ the standard deviation). (b) Acceleration (using the wheel sensor) and tension (using the force sensor) measured in the setup of the modified Atwood machine. (c) A group of students carrying out the experiment with the modified Atwood machine. }
\end{figure}

In this module, focused on Newton's Laws for the 11th grade, students begin with an initial observation of dynamic friction by sliding the iOLab device across a desk and analyzing the acceleration data captured. Through this activity, they are encouraged to explore instants in which the device reach is maximum speed and when the applied force ceases. Subsequently, they test their explanations using the force sensor, and from the graph, they can check the instant when the force exerted by the hand is no longer in contact with the device, allowing them to reject alternative explanations.  Having observed that that the horizontal segment of the graph (see Figure \ref{fig:mechanics} (a)) represents the phase in which dynamic friction is at work, students can then carry out an application experiment to estimate the friction coefficient in at least two different ways. This activity is inspired by the iOLab physics lab course created by the device’s developers \cite{iolab}. A similar activity has also been developed using smartphones \cite{Rakestraw2023}.

Following this, students further test their understanding by constructing a horizontal Atwood machine on their desks, both without and with a pulley (see Figures \ref{fig:mechanics} (b) and (c)). This challenges them to compare the experimental results with the mathematical model previously studied, taking into account the friction at the edge of the table, and refining their process.
The module then moves on to circular motion, where students are required to hold the iOLab in their hands with arms fully extended and then rotate using their body. They collect data and test the model of radial acceleration, $a = \omega^2 r$ by using a spreadsheet.  

\subsection{DC circuit}
\begin{wrapfigure}{l}{0.5\textwidth}
    \centering
    \includegraphics[width=0.48\textwidth]{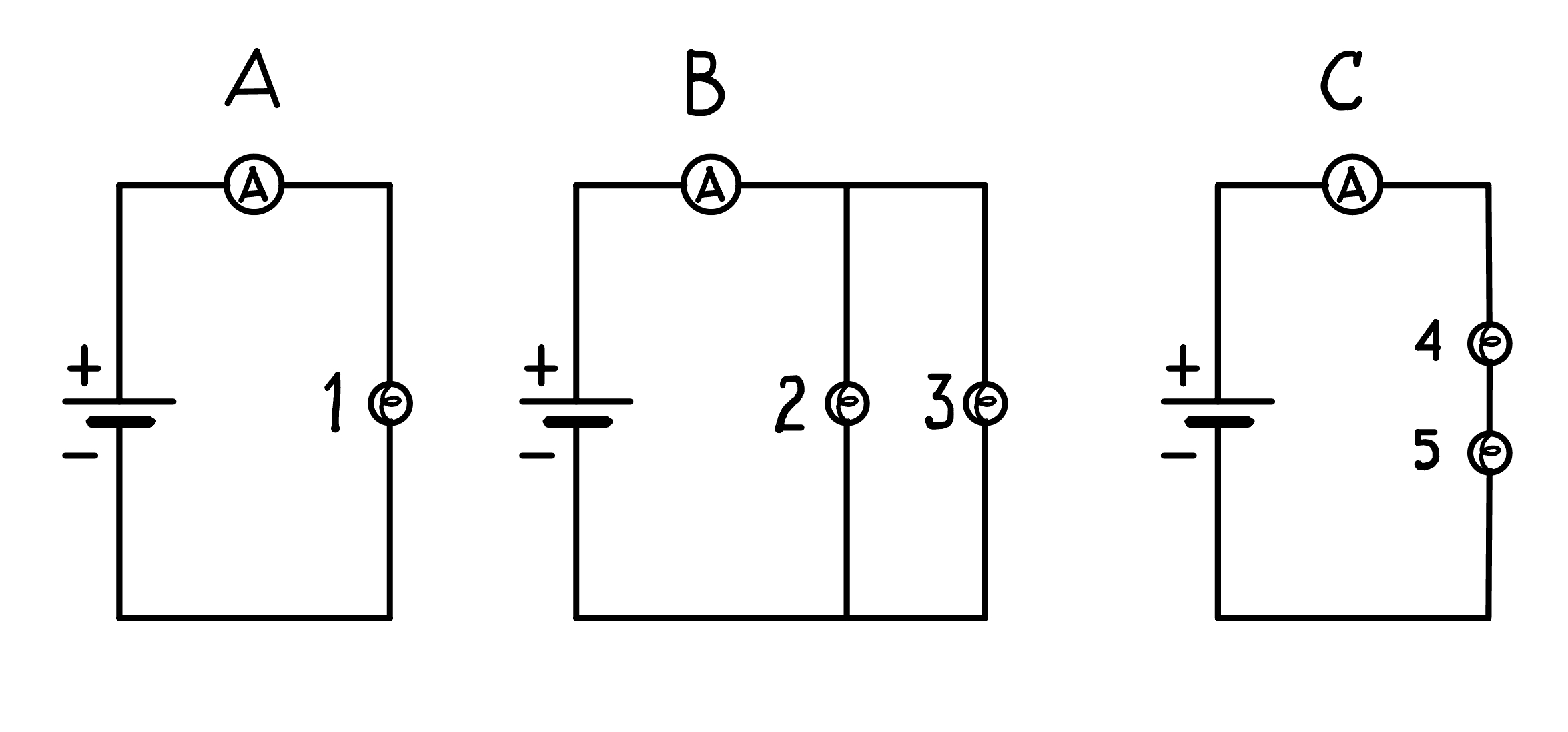}
    \caption{\label{fig:circuit} This figure represents the ranking exercise provided by the textbook \cite{Etkina2019PEarsonBook}. Three circuits with identical bulbs and emf sources are shown. The task required students to rank the circuits based on the ammeter readings, from the largest current to the smallest.}
\end{wrapfigure}

In this ISLE-based module, proposed in three  11th grade classrooms (see here \cite{Etkina2019IOPbook} for more details), students compare the behaviour of incandescent bulbs and LEDs in a simple circuit. 
The activity begins with an observational experiment in which students are asked to light both an LED and an incandescent bulb (separately) using basic circuit components, investigating how component orientation affects circuit functionality. The activity moves on to testing their explanations. They have to revise their explanations in response to unexpected results from one of the testing experiments (the LED glows but the incandescent bulb in series does not), thus fostering a deeper understanding of circuit behaviour and the iterative nature of the scientific method. After the ISLE-based activity, students engage in group discussions on conceptual circuit issues that have been documented in physics education research as potentially challenging such as those from McDermott \cite{McDermott1992} and considered in research-based assessments \cite{Madsen2021}. This approach is designed to help students to reflect and build on their difficulties. For example, the Figure \ref{fig:circuit} (adapted from \cite{Etkina2019PEarsonBook}) illustrates an exercise from a group activity in which students worked collaboratively using whiteboards to rank a set of circuits based on ammeter readings, with the largest ammeter reading listed first.The task promoted a deeper understanding of the underlying physical concepts related to current and resistance.

\begin{figure}[tbhp]
\begin{center}
\includegraphics[width=4.8in]{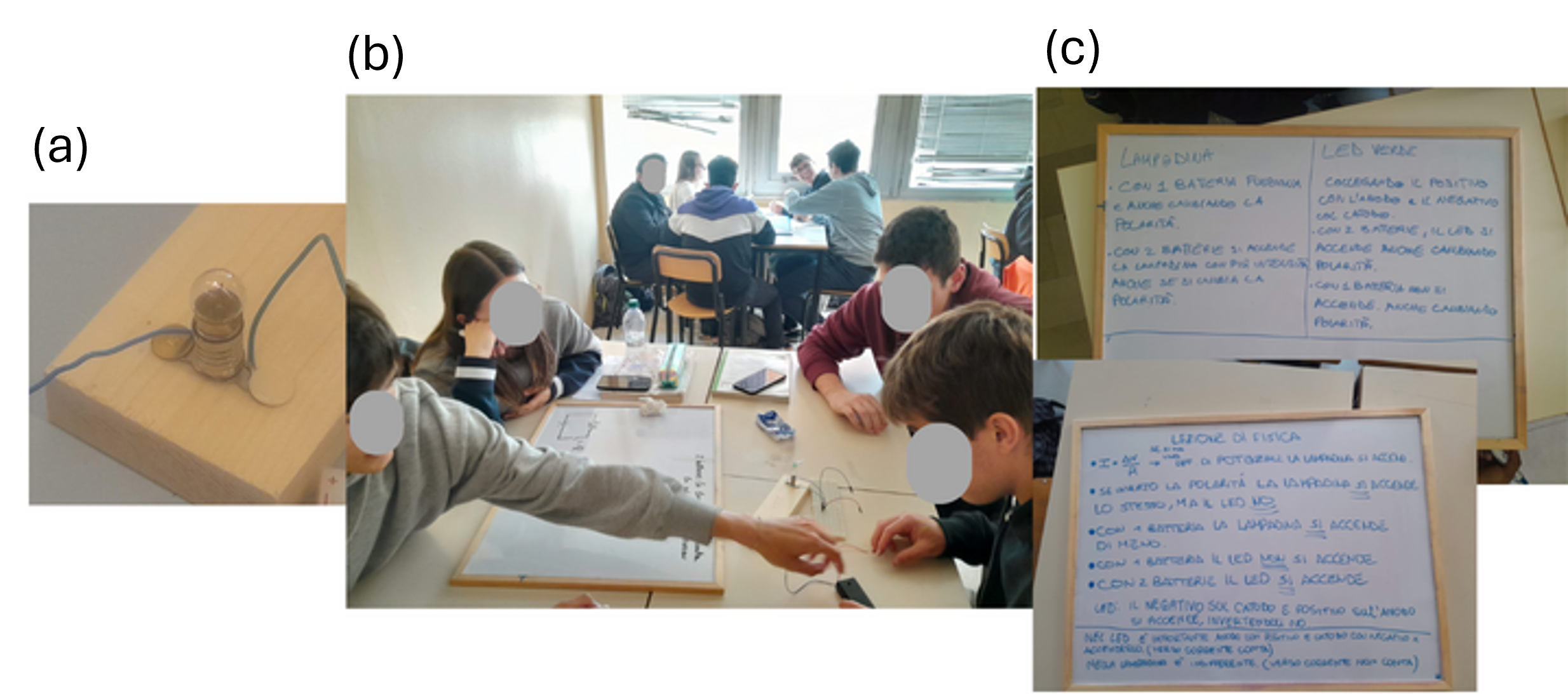}
\end{center}
\caption{\label{fig:LED} (a) An incandescent bulb mounted on a customised wooden stand for student experiments. (b) Students actively involved in the experimental task, exploring the basics of DC circuits. (c) Examples of student explanations and results presented on whiteboards after the investigation.}
\end{figure}

They then investigate Ohm's law and the I-V characteristic curves of resistors and lightbulbs using a PhET simulation and data tables. They recognize that while the resistance of resistors remains constant, the resistance of incandescent bulbs varies with the current. Subsequently, the focus shifts to constructing a setup with the IOLab system to examine the I-V characteristics of an LED. This involve placing a known resistor in series to the LED, measuring the potential difference across its terminals and calculating the current. Through the process of building the circuit with the IOLab, making measurements and collecting data, students engage in a comprehensive experimental investigation of electrical circuits (see Figure \ref{fig:LED}).

In analyzing the data collected, students are confronted with the asymmetrical nature of the I-$\Delta V$ curve for LEDs, in comparison to the symmetrical curve observed for incandescent lamps. This discovery is significant because it highlights the directional behavior of LEDs in circuits, which contrasts with the non-directional nature of incandescent bulbs. In addition, the students discover that LEDs have a distinct  threshold voltage which links back to their earlier qualitative investigations.

\section{Methods and data collection}
\begin{figure}
\begin{center}
\includegraphics[width=4.5in]{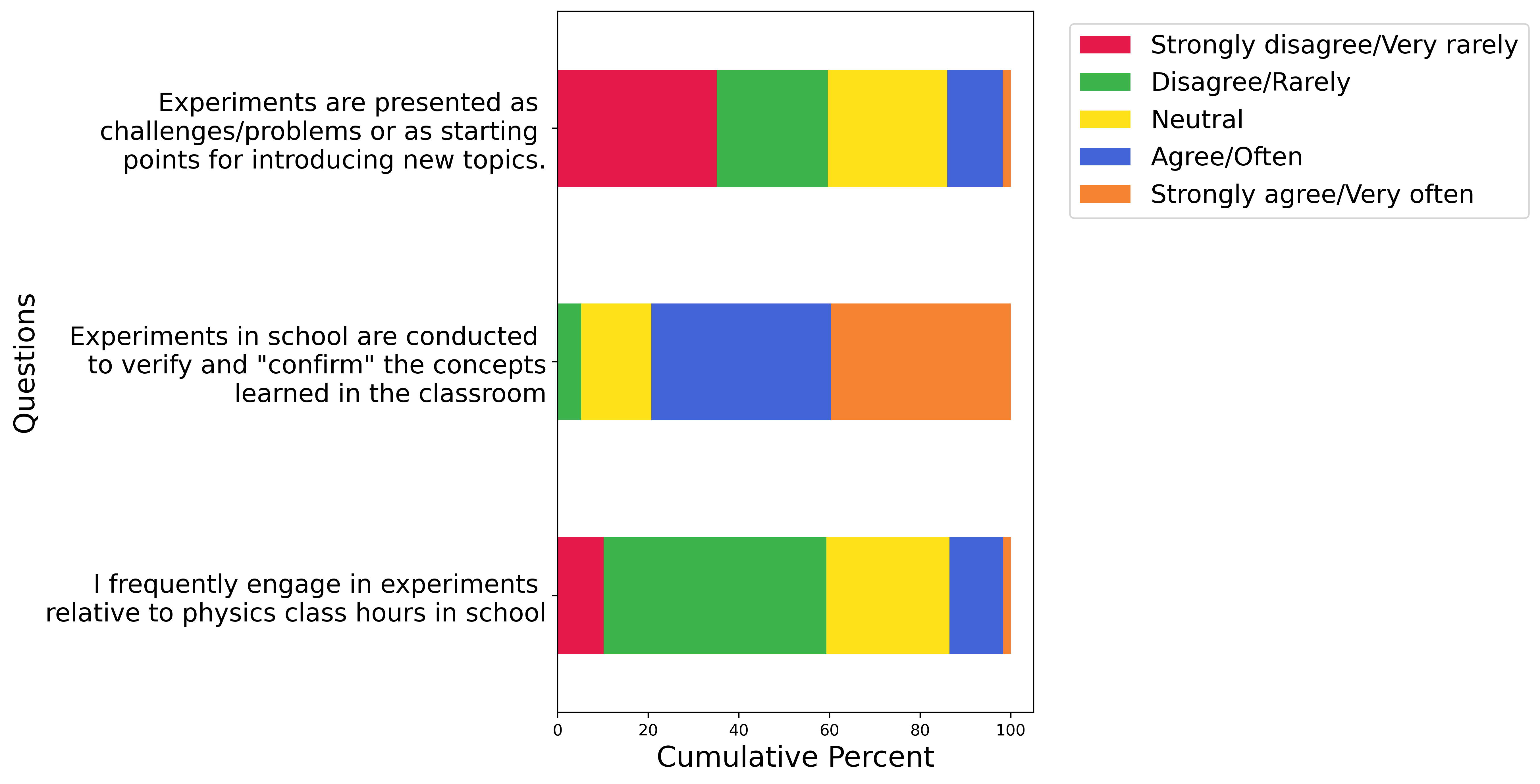}
\end{center}
\caption{\label{fig:presurveyhighschool} Stacked bar plots representing students'responses of all classrooms to three key questions of the pre-survey.}
\end{figure}

To evaluate the impact of our educational modules on student engagement and learning in physics experimental activities, we implemented a comprehensive evaluation strategy that combined both qualitative and quantitative research methods. Firstly, an initial survey was distributed to assess students' previous experience of working in the lab, their attitudes towards physics lab activities and their expectations. In Figure \ref{fig:presurveyhighschool} we report the cumulative percentages for three questions across three classrooms of the 11th grade and two classrooms of the 12th grade (see Table \ref{tab:overview}). The pre- and post-surveys, along with the final assessments, were individually completed by the students. In the post survey we received 44 responses from the two 12th grade classrooms and 56 responses from the three 11th grade classrooms.
In parallel, we analyzed students' outputs, which included short lab reports, collaborative tasks documented on classroom whiteboards, and a final test to assess the learning  objectives previously described. This test, with a mix of multiple choice, true/false and open-ended questions, was designed to assess students' conceptual knowledge, technical skills acquired using the iOLab apparatus and their understanding of the ISLE process. 
The test for the 12th grade class and lab resources are available on GitHub \cite{GithubRep}.
This assessment was complemented by a post-module survey to gather student feedback on the lab experience. 
The modules were implemented in two 11th grade classes at the "Liceo Scientifico opzione scienze applicate" lasted almost 10 hours of teaching time, about twice as long as the other modules. This extended timeframe allowed for the integration of an additional assessment part, where students were asked to self-assess their perceived proficiency before and after the module in a carefully selected subset  of scientific abilities \cite{Etkina2006Abilities}, which were chosen for their direct relevance to the activities carried out within the module.
\section{Results and discussion}
The primary objective of the evaluation component of this study was to determine the impact of integrating ISLE methodology with iOLab digital devices on improving student engagement, understanding, and skill development in physics laboratories. The following sections detail the results of this integration, highlighting the observed improvements in student engagement and development of scientific skills.

\subsection{Student Perception of the Lab experience}
To assess student engagement with the lab modules, we developed a questionnaire following general guidelines, such as those outlined by \cite{Lovelace2013}, and further tailored by us. This questionnaire was designed to gather feedback on the perceived usefulness of the lab activities and the questions are available at this link on GitHub \cite{GithubRep}. We summarize the results related to the students' attitudes towards the introduced modules, primarily utilizing Likert scale questions ranging from 1 to 5. In response to the question "2) The activities proposed were fun and interesting", out of 100 students (the students of all classrooms), 61 rated them highly (61\%), 33 were neutral (33\%) and 6 gave lower ratings (6\%).
This indicates a largely positive reception with some room for improvement.
Regarding the capability to use iOLab or similar sensor devices for experiments, from 100 students, 43 (43\%) felt confident ('4' or '5' rating), 29 (29\%) were neutral, and 19 (19\%) expressed lower confidence ('1' or '2'). For the 12th grade classes, which had more hours available and engaged in group activities using whiteboards, 35 students rated the experience positively ('4' or '5'), 8 were neutral, and only 1 student gave a low rating ('1'). This positive attitude is essential, as  students' epistemic beliefs about a subject are known to significantly influence how they learn \cite{Lising}.
From open-ended question analysis regarding the most appreciated aspects, it is evident that students particularly valued the use of whiteboards (12th grade classrooms) and group work for both exercises and experiments. Many students highlighted the importance of applying theoretical knowledge in practice. Additionally, learning technical aspects, such as the use of iOLab and spreadsheet software, was also emphasized. Students found the process of experimenting, then comparing collected data with hypotheses, to be particularly useful. In terms of problematic aspects, many students reported no issues, indicating a positive reception of the module. The most common feedback was a desire for more time to conduct experiments and a preference  for theoretical explanations before activities, aligning with students’ familiarity with traditional learning approaches.

\subsection{Learning Goals Assessments}
\begin{figure}
\begin{center}
\includegraphics[width=\textwidth]{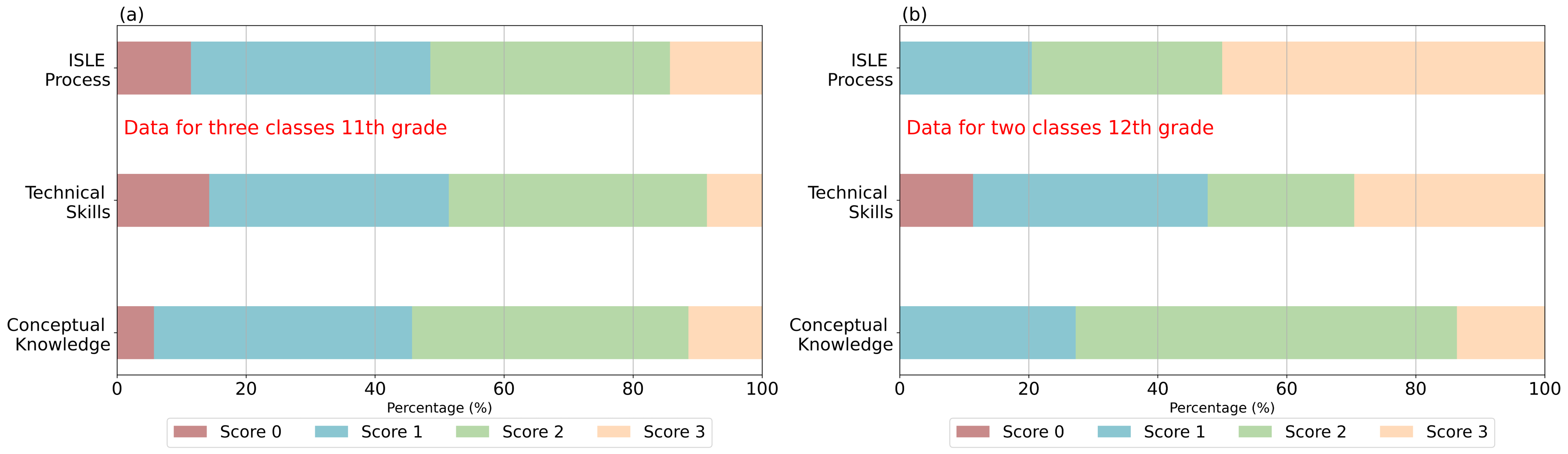}
\end{center}
\caption{\label{fig:LO} Percentage distribution of students' proficiency levels across three assessment dimensions—'ISLE process,' 'Conceptual knowledge,' and 'Technical skills'—as evaluated in the final test. Figure (a) shows results for three 11th grade classes (N=56), and Figure (b) for two 12th grade classes (N=44). Proficiency levels are categorized as 0 (missing), 1 (inadequate), 2 (needs improvement), and 3 (near mastery).}
\end{figure}

This section looks at the assessment of students' skills based on the final test results. We used a descriptive rubric for this analysis, focusing on three dimensions: Conceptual Knowledge, Technical Skills and the ISLE process. The rubric delineates four distinct levels of proficiency: 'Missing', 'Inadequate', 'Needs Some Improvement' and 'Near Mastery'. In Figure \ref{fig:LO} we present the assessment results, showing the proportion of students achieving each level of proficiency. Part (a) presents the data from the three grade 11 classrooms, while part (b) details the results from the two 12th grade classrooms. The results indicate a positive trend, with the majority of students demonstrating good levels of Technical Skills, Conceptual Knowledge, and familiarity with the ISLE Process. Notably, the outcomes are slightly better for the two 12th grade classes, where a greater number of instructional hours were allocated. This suggests that the extended engagement in these areas influences student proficiency.
To further investigate the two 12th grade classes, students were asked in the final test to self-assess their proficiency in the scientific abilities (or sub-abilities) that we considered to be central to the module, using the same 5-point Likert scale. As highlighted by previous research \cite{Etkina2008}, the development of these abilities typically takes about 8 weeks for college students and benefits from continuous feedback through rubrics used by both students and teachers. Despite these relevant limitations, our aim was to explore students' perceptions of specific abilities  used during the module.
The most appropriate way to present Likert scale responses would be to use histograms to visually represent the distribution of Likert scale responses. However, due to space limitations, we present here the results in tabular form, summarizing the mean, standard error and effect size (Cohen's d) of the responses (see Table \ref{tab:skills_assessment}).

\begin{table}[h!]
\centering
\caption{Comparison of pre- and post-self-assessment on students' scientific sub-abilities (e.g., "Being able to...") for the 12th grade (N=44), including average scores, standard error (SE), and effect size.}
\begin{tabular}{p{5.6cm}ccc}
\hline
\textbf{Question} & \textbf{Pre (Mean $\pm$ SE)} & \textbf{Post (Mean $\pm$ SE)} & \textbf{Effect Size} \\
\hline
Describe what was observed & 2.8 $\pm$ 0.1 & 3.7 $\pm$ 0.1 & 1.24 \\
Develop an explanation based on observations & 2.7 $\pm$ 0.1 & 3.5 $\pm$ 0.1 & 0.94 \\
Design an experiment to observe a phenomenon & 2.5 $\pm$ 0.2 & 3.4 $\pm$ 0.1 & 0.97 \\
Predict experiment results based on theoretical hypotheses & 2.9 $\pm$ 0.2 & 3.5 $\pm$ 0.1 & 0.59 \\
Comment on the agreement/disagreement of data with predictions & 3.0 $\pm$ 0.2 & 3.7 $\pm$ 0.1 & 0.78 \\
Identify hypotheses or assumptions in an experiment & 2.7 $\pm$ 0.1 & 3.4 $\pm$ 0.1 & 0.85 \\
Record and represent data significantly & 3.1 $\pm$ 0.1 & 3.8 $\pm$ 0.1 & 0.83 \\
Analyze data appropriately in an experiment & 3.0 $\pm$ 0.1 & 3.7 $\pm$ 0.1 & 0.83 \\
\hline
\end{tabular}

\label{tab:skills_assessment}
\end{table}
The statistical analysis using the Wilcoxon test showed significant improvements in students' self-assessment of all 8 scientific (sub-) abilities assessed from before to after the teaching intervention. Each question had a p-value below the significance threshold ($\alpha = 0.05$), confirming significant changes in students' perceptions. 
In addition to statistical significance, the effect size (Cohen's d) also indicates the practical significance of these changes, with values ranging from moderate to large across all abilities. Specifically, effect sizes ranged from 0.59 to 1.24, indicating significant improvements in students' perceived confidence and ability in science. The largest effect size (1.24) was observed in students' ability to describe observed phenomena, highlighting the strong impact of the intervention on this skill.
The quantitative results, combined with the increase in mean scores from pre to post and the reduced standard errors, indicate an increase in students' perceived confidence in scientific abilities. While these results suggest a positive trend, we recognize the limitations of student self-assessment, as it may not always reflect objective progress. 
Educational psychologists have extensively explored how individual perceptions and beliefs impact motivation and academic performance (for example \cite{Zimmerman2000}). These studies indicate that expectancy components play a critical role in shaping how students engage with learning tasks and persevere through challenges. Bandura's concept of self-efficacy—the belief in one's own ability to succeed in specific situations—further explains this link, showing that students with a strong sense of self-efficacy are more likely to adopt effective learning strategies, set higher goals, and maintain resilience when faced with academic obstacles \cite{Bandura1986}. Given its impact on learning outcomes, self-efficacy remains an important factor to explore in educational contexts. The findings from our study, suggesting positive student engagement and skill development through the ISLE methodology integrated with iOLab devices, align with the observations made by Rutberg et al. \cite{Rutberg2019}. Their investigation into the impact of ISLE-based labs in courses maintaining traditional lecture formats revealed that, while the growth in student scientific abilities was not as pronounced as in fully integrated ISLE environments, significant improvements were still observed.

\section{Conclusions}
In our study, we integrated the ISLE approach with iOLab digital devices to promote an active learning environment in several Italian high school physics classrooms. This innovative strategy was designed to increase student engagement and improve understanding through hands-on experimentation and data analysis. While our preliminary results do not allow for a direct comparison with traditional teaching methods, due to the absence of a control group—primarily because no parallel classes were available that followed the same learning goals using traditional methods, they suggest significant pedagogical benefits from incorporating technology into laboratory settings, improving both the learning experience and student outcomes.
The study’s main limitations include its brief duration and the separation of labs from theoretical lessons.
Conducted over a few hours, the modules may not fully demonstrate the ISLE approach or the potential of the iOLab device to influence long-term learning and skill development. In addition, isolating laboratory activities from related lecture content may weaken the pedagogical coherence and cumulative impact of integrating theory and practice.

Nevertheless, the results were positive, particularly in classrooms where the intervention lasted longer. Students quickly became familiar with the iOLab equipment within a few sessions. 
The intervention received positive feedback, with students appreciating the importance of group work and hands-on activities using technology. A particular challenge was identified during the dynamic friction activity, where students struggled to interpret acceleration graphs. Addressing such challenges in lectures throughout the course, for example by highlighting force and motion diagrams for different scenarios, could potentially alleviate these problems and help students construct their knowledge. This strategy highlights the value of using identified challenges as a basis for deeper exploration and understanding in future lessons.

Nevertheless, the ISLE process requires more time to realize its full benefits. The limited timeframe of our study suggests that future research should adopt a comprehensive approach that more closely integrates theoretical and practical components. In addition, a longer study of at least one year would provide a better opportunity to test the ISLE approach in non-US contexts and to address the challenges that may arise, particularly in integrating the methodology into different educational systems and cultural environments.
In work that has followed on from that described in this paper, one of the authors, as a high school teacher, was able to apply the ISLE methodology more fully, including its assessment practices and the use of supporting materials such as textbooks and exercises. This experience supported the potential of ISLE to promote a deeper understanding of physics, given sufficient time and resources. The results of this research are preliminary and merit further investigation. Our experience indicates that the integration of innovative technologies such as iOLab with the ISLE approach is feasible and can enhance learning and student engagement at the high school level. While the results are promising, 
further studies are needed to confirm the long-term impact of this combination on students' understanding of physics concepts and their active participation in the learning process.
\ack{}
We would like to thank the teachers and the students of the Italian schools for allowing us to carry out our intervention and for their hospitality.We are also grateful to the reviewers and the editors for their constructive comments, which have contributed to improving the quality of this work.

\end{document}